\documentclass{article}

\usepackage{PRIMEarxiv}
\usepackage{natbib}
\usepackage[utf8]{inputenc} 
\usepackage[T1]{fontenc}    
\usepackage{hyperref}       
\usepackage{url}            
\usepackage{booktabs}       
\usepackage{amsfonts}       
\usepackage{nicefrac}       
\usepackage{microtype}      
\usepackage{lipsum}
\usepackage{fancyhdr}       
\usepackage{graphicx}       
\graphicspath{{media/}}     
\usepackage{amssymb}
\usepackage{amsmath}

\usepackage{url}
\usepackage{lineno}
\usepackage{amsmath, amsfonts, amssymb, amsthm}
\usepackage[table]{xcolor}
\usepackage{booktabs}
\usepackage{longtable}
\usepackage{colortbl}
\usepackage{tikz}
\usetikzlibrary{arrows.meta,positioning,calc,fit,shadows.blur}

\usepackage[utf8]{inputenc}   
\usepackage[T1]{fontenc}

\usepackage{caption}
\definecolor{ParametricBlue}{RGB}{33,114,194}   
\definecolor{NonparamGreen}{RGB}{46,139,87}     
\definecolor{BayesOrange}{RGB}{230,135,0}       
\definecolor{DLRed}{RGB}{200,50,50}             
\definecolor{TierGray}{gray}{0.15}

\tikzset{
  box/.style={draw, rounded corners=2pt, align=center,
              minimum height=8mm, inner xsep=4mm, inner ysep=3mm,
              font=\footnotesize, white, text width=30mm},
  catbox/.style={box, text width=37mm, minimum height=9mm, font=\small\bfseries},
  lane/.style={draw=black!10, fill=black!2, rounded corners=3pt},
  arrow/.style={-{Latex}, thick},
  dasharrow/.style={-{Latex}, thick, dashed},
  tierlabel/.style={black!70, font=\bfseries\small}
}

\usepackage{xcolor}
\newcommand{\tr}[1]{\textcolor{black}{#1}}

\title{Efficacy Analysis in Clinical Trials: A Comprehensive Review of Statistical and Machine Learning Approaches
}

\author{
  Dhrubajyoti Ghosh \\
  Department of Biostatistics and Bioinformatics \\
  Duke University \\
  Durham\\
  \texttt{dg302@duke.edu} \\
   \And
  Samhita Pal \\
  Department of Statistics \\
  North Carolina State University \\
  Raleigh\\
  \texttt{spal4@ncsu.edu} \\
}

\begin{document}
\maketitle

\begin{abstract}
  Efficacy testing is a cornerstone of clinical trials, ensuring that medical interventions achieve their intended therapeutic effects. Over the decades, a wide range of statistical methodologies have been developed to address the complexities of clinical trial data, including parametric, nonparametric, Bayesian, and machine learning approaches. Parametric methods, such as t-tests, ANOVA, and LMMs, have traditionally been the foundation of efficacy testing due to their efficiency under well-defined assumptions. Nonparametric techniques, including the Friedman test, Brunner-Munzel test, and modern extensions like nparLD, have emerged as robust alternatives, particularly for skewed, ordinal, or non-normal data. Bayesian methodologies have enabled the incorporation of prior information and uncertainty quantification, while machine learning techniques, such as deep learning and reinforcement learning, are revolutionizing trial designs and outcome predictions. Despite these advancements, significant gaps remain, including challenges in handling high-dimensional data, missingness, and ensuring equitable efficacy testing across diverse populations. This review provides a comprehensive overview of these statistical methods, highlighting their applications, strengths, limitations, and future directions. By bridging traditional statistical frameworks with modern computational techniques, the field can continue to advance toward more reliable and personalized clinical trial methodologies.
\end{abstract}

\keywords{Efficacy testing \and longitudinal \and cross-sectional \and clinical trials \and parametric methods \and non-parametric methods \and Bayesian methods \and machine learning \and deep learning.}

\section{INTRODUCTION}

Efficacy testing is a fundamental aspect of clinical trials, assessing whether a medical intervention produces the intended therapeutic effect under ideal or controlled conditions. It serves as the cornerstone for determining the effectiveness of drugs, treatments, and devices, guiding regulatory approvals and informing clinical practice. Statistical methods are pivotal in this process, ensuring that efficacy conclusions are robust, reproducible, and scientifically valid. Over the years, advancements in statistical methodologies have enabled researchers to tackle the complexities inherent in efficacy testing, particularly as datasets have become more intricate and multidimensional.

Parametric methods have traditionally dominated efficacy testing in clinical trials due to their simplicity and statistical efficiency. Classical approaches, such as t-tests and analysis of variance (ANOVA), remain staples for comparing treatment effects across groups \citep{student1908probable, fisher1925}. For longitudinal and repeated measures data, linear mixed-effects models (LMMs) have become the method of choice, enabling the modeling of individual patient trajectories while accounting for variability across and within subjects \citep{laird1982, verbeke2000}. However, parametric methods rely heavily on assumptions of normality and homogeneity of variance, which are often violated in real-world clinical trials, particularly with heterogeneous or skewed data.

To address these challenges, nonparametric methods have emerged as robust alternatives for efficacy testing. These methods, including the Friedman test \citep{friedman1937}, Brunner-Munzel test \citep{brunner2000}, and modern tools like nparLD \citep{noguchi2012}, provide greater flexibility by relaxing distributional assumptions. Nonparametric approaches are particularly suited for ordinal outcomes, non-normal data, and small sample sizes, making them invaluable in certain therapeutic areas and rare disease studies. Additionally, recent advancements, such as the Longitudinal Rank Sum Test (LRST) \citep{xu2025novel, ghosh2025non}, have further extended the utility of nonparametric methods to longitudinal settings, enabling the analysis of time-dependent efficacy outcomes.

Bayesian methods have also gained prominence in efficacy testing by allowing researchers to incorporate prior knowledge and account for uncertainty in treatment effects. Hierarchical Bayesian models enable the pooling of information across subgroups, while Bayesian nonparametric approaches, such as Gaussian processes, provide flexible tools for modeling complex, nonlinear relationships in longitudinal data \citep{gelman2013, rasmussen2006}. These methods are particularly advantageous in adaptive trial designs, where interim results inform subsequent decisions, and in personalized medicine, where individual-level predictions are critical.

In recent years, machine learning (ML) techniques have revolutionized efficacy testing by addressing challenges associated with high-dimensional, unstructured, and multimodal datasets. Methods such as deep learning, reinforcement learning, and federated learning have been applied to optimize trial designs \citep{gligorijevic2019optimizing}, predict treatment outcomes \citep{zhu2023prediction}, and improve dosing regimens \citep{nemati2016optimal}. Despite their promise, these approaches face challenges related to interpretability, generalizability, and integration with traditional statistical frameworks.

\tr{The choice of analytical framework in efficacy testing depends on the underlying research objective and the structure of the clinical data. Broadly, these objectives can be grouped into four categories: estimating and comparing treatment effects under defined assumptions, assessing relative efficacy through distributional comparisons with minimal assumptions, quantifying uncertainty while incorporating prior knowledge, and predicting individual-level outcomes or response patterns in complex data. Parametric methods are most appropriate when the goal is to estimate average treatment differences or model temporal trajectories under well-specified distributional assumptions. Nonparametric approaches are valuable when the data are skewed, ordinal, or heterogeneous, allowing robust comparisons that remain valid under weaker conditions. Bayesian models provide a flexible probabilistic framework that integrates prior evidence with observed data, supporting hierarchical and adaptive trial designs. Finally, machine learning techniques, ranging from deep neural architectures to reinforcement learning, are designed for predictive and exploratory analyses, capable of capturing complex, nonlinear dependencies in high-dimensional or multimodal datasets. By explicitly linking these analytical families to their corresponding research objectives, this review aims to guide researchers in selecting suitable methodologies for efficacy analysis across diverse clinical settings.}

\tr{Missing data are almost unavoidable in clinical and biomedical research, and the assumptions made about how data become missing have direct consequences for statistical validity. The standard taxonomy distinguishes among three mechanisms \citep{little2019statistical, molenberghs2007missing}. Data are missing completely at random (MCAR) when the probability of missingness is unrelated to both observed and unobserved outcomes; complete-case analysis remains unbiased but inefficient in this setting. Data are missing at random (MAR) when missingness depends only on observed variables; likelihood-based estimation and multiple imputation can then provide valid inference if the model for the observed data is correctly specified. Data are missing not at random (MNAR) when missingness depends on unobserved outcomes, leading to potential bias even under correct modeling of the observed data. In such cases, specialized models such as selection models, pattern-mixture models, or shared-parameter frameworks must explicitly characterize the missingness mechanism \citep{daniels2008}. Because these assumptions underpin all major statistical paradigms, the subsequent sections highlight how each methodological family handles missingness under these three mechanisms.}

This review provides a comprehensive examination of statistical methods for efficacy testing in clinical trials, categorizing them into parametric, nonparametric, Bayesian, and machine learning approaches. By highlighting the strengths, limitations, and applications of these methodologies, the paper aims to guide researchers in selecting appropriate tools for efficacy analysis in diverse clinical settings. Furthermore, the review explores recent methodological advancements, emphasizing their potential to address emerging challenges in clinical trial research. We have explored parametric tests in Section \ref{sec:parametric}, nonparametric tests in Section \ref{sec:nonparametric}, Bayesian methods in Section \ref{sec:bayesian} and some popular Deep Learning based methodologies in Section \ref{sec:deepLearn}.

\section{PARAMETRIC TESTS}
\label{sec:parametric}

Parametric tests are among the most widely used statistical methods in clinical trials, owing to their simplicity, efficiency, and strong inferential properties under well-defined assumptions. \tr{Parametric approaches are most suitable when the objective is to estimate mean treatment differences or model trajectories under defined distributional assumptions.} These methods rely on specific distributional assumptions, such as normality of the data and homogeneity of variances, making them particularly suited for continuous and normally distributed outcomes. Commonly applied parametric tests include the t-test, analysis of variance (ANOVA), and analysis of covariance (ANCOVA), each tailored for comparing means across groups or conditions under varying experimental designs. In many clinical trials, efficacy is evaluated based on the change from baseline to the final endpoint, where cross-sectional parametric tests like the t-test or ANCOVA can be used to compare treatment groups \citep{vickers2001, pocock2002}. For longitudinal or repeated measures data, LMMs extend the scope of parametric analysis by incorporating random effects to account for within-subject correlations and handling missing data under the missing-at-random (MAR) framework \citep{laird1982, verbeke2000}. While parametric methods are robust under their assumptions, their utility can be limited when these assumptions are violated, such as in the presence of skewed distributions or heterogeneous variances, highlighting the need for alternative approaches in certain clinical trial settings.

\subsection{Cross-Sectional Parametric Tests}

Efficacy testing in clinical trials often begins with the comparison of treatment groups at a single time point, where parametric methods provide powerful and interpretable statistical tools. These methods, including t-tests, analysis of variance (ANOVA), and analysis of covariance (ANCOVA), leverage assumptions about data distribution to efficiently estimate treatment effects. The following section discusses these cross-sectional parametric tests, their applications in clinical research, and their advantages and limitations in different trial settings.

\subsubsection{t-Tests}

The t-test, introduced by William Sealy Gosset in 1908 under the pseudonym “Student,” is a parametric statistical test designed to compare the means of two groups under specific assumptions \citep{student1908probable}. The standard t-test, also called the independent t-test, evaluates the difference between two independent group means. The test statistic is calculated as:
\(
t = \left(\bar{X}_1 - \bar{X}_2\right)/\sqrt{s_p^2 ( n_1^{-1} + n_2^{-1})},
\)
where \( \bar{X}_1 \) and \( \bar{X}_2 \) are the sample means, \( n_1 \) and \( n_2 \) are the sample sizes, and
\(
s_p^2 = \left((n_1 - 1)s_1^2 + (n_2 - 1)s_2^2\right)/\left(n_1 + n_2 - 2\right)
\)
is the pooled variance. Here, \( s_1^2 \) and \( s_2^2 \) represent the variances of the two groups. The paired t-test, a variation of the standard t-test, is used for dependent or matched samples, analyzing the mean differences between paired observations using the formula:
\(
t = \bar{d}/\left(s_d / \sqrt{n}\right),
\)
where \( \bar{d} \) is the mean of the differences, \( s_d \) is the standard deviation of the differences, and \( n \) is the number of paired observations. Another important extension, Welch’s t-test, modifies the standard t-test to account for unequal variances and sample sizes between groups, with its test statistic computed as:
\(
t = \left(\bar{X}_1 - \bar{X}_2\right)/\sqrt{({s_1^2}/{n_1}) + ({s_2^2}/{n_2})}.
\)
These methods, which rely on the assumption of normally distributed data, have become foundational tools for statistical inference in clinical trials and experimental research \citep{welch1947generalization}.

\begin{longtable}[l]{p{2.5cm}p{9.5cm}p{4.5cm}}
\caption{Applications of the t-Tests in Clinical Trials \label{tab:ttest}} \\
\hline
\rowcolor{gray!20} \textbf{Study} & \textbf{Application} & \textbf{Therapeutic Area} \\
\endfirsthead

\multicolumn{3}{c}%
{{\bfseries Table \thetable\ Continued from previous page}} \\
\hline
\rowcolor{gray!20} \textbf{Study} & \textbf{Application} & \textbf{Therapeutic Area} \\
\endhead

\hline
\endfoot

\hline
\endlastfoot

\cite{berger1987testing} & Evaluation of pre- and post-treatment differences in blood pressure levels & Cardiovascular Research \\
\rowcolor{gray!10} 
\cite{psaty1997health} & Assessment of the impact of antihypertensive therapies on systolic and diastolic blood pressure in hypertensive patients & Hypertension \\
\cite{garber1997efficacy} & Analysis of Metformin's impact on glycemic control using independent t-tests to compare pre- and post-intervention HbA1c levels & Diabetes \\
\rowcolor{gray!10} 
\cite{fogelholm2000does} & Analysis of the efficacy of dietary and behavioral therapies in weight-loss interventions & Weight-Loss Interventions \\
\cite{zimmerman2004} & Evaluation of psychometric properties in mental health research & Mental Health \\ 
\rowcolor{gray!10} 
\cite{vinkers200415} & Evaluation of cognitive behavioral therapy's impact on depression scores in mental health interventions & Mental Health \\
\cite{ferrario2005effect} & Examination of the effects of angiotensin-converting enzyme inhibitors on cardiac function & Cardiology \\
\rowcolor{gray!10} 
\cite{ruxton2006} & Comparison of drug efficacy for treatments with differing variance profiles & General Healthcare \\ 
\cite{sousa2009systematic} & Systematic reviews to synthesize data on pre- and post-treatment changes in diagnostic accuracy studies & Meta-analytical Frameworks \\
\rowcolor{gray!10} 
\cite{fagerland2011} & Application of Welch’s t-test in contingency table analyses for robustness in healthcare data & Healthcare \\ 
\cite{rasch2011} & Exploration of Welch’s t-test in pharmacokinetics to address unequal variances due to population heterogeneity & Pharmacokinetics \\ 
\rowcolor{gray!10} 
\cite{dewinter2013} & Highlighted effectiveness of Welch’s t-test for small-sample comparisons in clinical studies & Clinical Research \\ 
\cite{pi2015randomized} & Assessment of the efficacy of Liraglutide in reducing body weight through paired analysis & Weight Management \\ 
\rowcolor{gray!10} 
\cite{beasley2022evaluation} & Examination of Budesonide-Formoterol therapy in improving lung function parameters through paired t-tests & Pulmonology \\ 
\cite{sotoudeh2024systematic} & Optimization of photon-counting CT for lung density quantifications using paired analysis & Radiology \\ 
\end{longtable}

T-tests are widely recognized for their simplicity, computational efficiency, and effectiveness in comparing means, particularly in small to moderate-sized datasets (Cf. Table \ref{tab:ttest}). The paired t-test enhances statistical power by focusing on within-subject differences, while Welch’s t-test provides a robust alternative for scenarios involving unequal variances or sample sizes. However, these methods rely on assumptions of normality and are sensitive to outliers, which can compromise their validity in non-normal or heavily skewed data. Additionally, t-tests are limited to mean-based comparisons, rendering them unsuitable for ordinal data or high-dimensional datasets that require more sophisticated statistical approaches, such as mixed-effects models or machine learning techniques.

\subsubsection{Variance Analysis Methods}
Analysis of Variance (ANOVA), introduced by Fisher in 1918, is a parametric method for comparing the means of multiple groups. It partitions total variation into between-group variation (\(SS_{\text{between}}\)) and within-group variation (\(SS_{\text{within}}\)), with the F-statistic calculated as \(
F = {SS_{\text{between}}(N - k)}/{SS_{\text{within}}(k - 1)}
\) where \(k\) is the number of groups and \(N\) is the total number of observations. Under the null hypothesis (\(H_0\)), the F-statistic follows an F-distribution with \(k-1\) and \(N-k\) degrees of freedom. ANOVA assumes normality within groups, homoscedasticity, and independence of observations. Extensions like repeated measures and two-way ANOVA have been developed for longitudinal and factorial designs. While ANOVA controls Type I error better than multiple pairwise t-tests, violations of its assumptions can lead to misleading results. Nonparametric alternatives like the Kruskal-Wallis test address these limitations. ANOVA is also less powerful with unequal group sizes.

Analysis of Covariance (ANCOVA), introduced by Fisher in 1925, combines ANOVA and linear regression to compare group means while adjusting for one or more covariates. ANCOVA is particularly useful in clinical trials to control for baseline imbalances (Cf. Table \ref{tab:anova}). The model is \(
Y_{ij} = \mu + \alpha_i + \beta X_{ij} + \epsilon_{ij}
\) where \(Y_{ij}\) is the outcome, \(X_{ij}\) is the covariate, and \(\beta\) is the regression coefficient for the covariate. ANCOVA adjusts the outcome by removing the effect of the covariate, enabling a comparison of group means for a common covariate value. It assumes linearity between the covariate and the outcome, homogeneity of regression slopes, and normality and homoscedasticity of residuals. ANCOVA has been extended to handle multiple covariates and more complex designs. However, violations of assumptions or failure to account for covariate-treatment interactions can lead to biased results.

\begin{longtable}[l]{p{2.5cm}p{2cm}p{7cm}p{4.5cm}}
\caption{Applications of ANOVA and ANCOVA in Clinical Trials}\label{tab:anova} \\
\hline
\rowcolor{gray!20} \textbf{Study} & \textbf{Method} & \textbf{Application} & \textbf{Therapeutic Area} \\
\endfirsthead

\multicolumn{4}{c}%
{{\bfseries Table \thetable\ Continued from previous page}} \\
\hline
\rowcolor{gray!20} \textbf{Study} & \textbf{Method} & \textbf{Application} & \textbf{Therapeutic Area} \\
\endhead

\hline
\endfoot

\hline
\endlastfoot

\cite{olson1976} & ANOVA & Multivariate studies of psychological interventions for chronic pain management & Chronic Pain Management \\
\rowcolor{gray!10} 
\cite{kay1977} & ANCOVA & Adjustment for tumor size at baseline in oncology trials to compare treatment efficacy & Oncology \\
\cite{pocock1987} & ANOVA & Comparison of the effects of beta-blockers across different dosages & Cardiovascular Research \\
\rowcolor{gray!10} 
\cite{wright1992} & ANCOVA & Longitudinal studies to address baseline imbalances in randomized controlled trials & Randomized Controlled Trials \\
\cite{senn1993} & ANOVA & Comparison of cognitive behavioral therapies across different durations of intervention & Mental Health \\
\rowcolor{gray!10} 
\cite{rogers199824} & ANCOVA & A 24-week, double-blind, placebo-controlled trial of donepezil in Alzheimer's disease & Alzheimer's Disease \\
\cite{pinheiro2000} & ANOVA & Extension to mixed models for longitudinal analyses in oncology trials & Oncology \\
\rowcolor{gray!10} 
\cite{vickers2001} & ANCOVA & Adjusting for percentage change from baseline in pain management studies, avoiding biased estimates & Pain Management \\
\cite{beaton2002} & ANOVA & Assessment of musculoskeletal burden differences across patient subgroups in observational studies & Musculoskeletal Disorders \\
\rowcolor{gray!10} 
\cite{pocock2002} & ANCOVA & Cardiovascular trials to adjust for baseline blood pressure and cholesterol levels & Cardiovascular Research \\
\cite{bradburn2003} & ANCOVA & Survival analysis to adjust for clinical staging in oncology trials & Oncology \\
\rowcolor{gray!10} 
\cite{maxwell2004} & ANOVA & Use of repeated measures to evaluate temporal effects of pain relief medications in drug trials & Pain Management \\
\cite{gueorguieva2004} & ANOVA & Analysis of repeated measures in antidepressant effects over time & Mental Health \\
\rowcolor{gray!10} 
\cite{senn2006} & ANOVA & Adjusting for random baseline imbalances in clinical trials & Clinical Trials \\
\cite{siddique2008} & ANCOVA & Handle missing data and apply covariate adjustment in weight-loss trials & Weight-Loss Trials \\ 
\rowcolor{gray!10} 
\cite{ghosh2025thanos} & ANCOVA & Incorporates social network structure and opinion for election forecasting & Election Prediction
\end{longtable}

\subsection{Longitudinal Parametric Tests}
While cross-sectional parametric tests such as t-tests and ANOVA are widely used in clinical trials, they are limited to analyzing treatment effects at a single time point. However, many clinical studies involve repeated measurements of patient outcomes over time, requiring statistical methods that account for within-subject correlations and time-dependent variations. Longitudinal parametric models, such as Linear Mixed-Effects Models (LMMs) and Generalized Estimating Equations (GEEs), extend traditional parametric frameworks to capture temporal dynamics and patient-specific variability. The following section explores these methods, highlighting their applications in efficacy testing and their ability to handle missing data and complex correlation structures.
\subsubsection{Linear Mixed Effects Model}


LMMs, introduced by \cite{henderson1954}, extend linear regression by incorporating both fixed and random effects, making them particularly suitable for analyzing hierarchical or longitudinal data, such as repeated measures in clinical trials. Fixed effects represent population-level relationships, while random effects capture subject-specific variability, allowing LMMs to effectively model within-subject or within-cluster correlations. The general model is expressed as:
\(
Y = X\beta + Zb + \epsilon,
\)
where \( X\beta \) denotes the fixed effects, \( Zb \) represents the random effects (\( b \sim N(0, G) \)), and \(\epsilon \sim N(0, R)\) accounts for residual errors. One of the significant strengths of LMMs is their ability to handle missing data under the Missing at Random (MAR) assumption, where the probability of missingness depends only on observed data. Unlike traditional methods that exclude incomplete cases, LMMs use all available data through maximum likelihood or restricted maximum likelihood estimation, reducing bias and improving efficiency. Additionally, LMMs accommodate complex covariance structures, such as autoregressive or unstructured correlations, providing flexibility in modeling dependencies.


LMMs, while versatile, have limitations. They assume linear relationships, normality of random effects and residuals, and independence between random and fixed effects. While diagnostic checks, transformations, and robust estimation techniques can address some violations, these assumptions inherently restrict their use in datasets with non-linear relationships or non-normal distributions. Computational challenges with large datasets or complex random-effects structures can be mitigated using efficient algorithms like Expectation-Maximization or parallel computing. However, LMMs struggle with data missing not at random (MNAR), as the mechanism depends on unobserved variables, and biased estimates may result when the independence of random and fixed effects is violated. Additionally, interpreting random effects in complex hierarchical models can be challenging, limiting subject-specific inferences. These limitations underscore the need to carefully evaluate assumptions and choose appropriate methods for specific research contexts.

\begin{longtable}[l]{p{2.5cm}p{2cm}p{7.5cm}p{4cm}}
\caption{Applications of LMMs in Clinical Trials} \\
\hline
\rowcolor{gray!20} \textbf{Study} & \textbf{Method} & \textbf{Application} & \textbf{Therapeutic Area} \\
\endfirsthead

\multicolumn{4}{c}%
{{\bfseries Table \thetable\ Continued from previous page}} \\
\hline
\rowcolor{gray!20} \textbf{Study} & \textbf{Method} & \textbf{Application} & \textbf{Therapeutic Area} \\
\endhead

\hline
\endfoot

\hline
\endlastfoot

\cite{laird1982} & LMM & Model patient responses over time, introducing random effects for patient-specific trajectories & Psychiatry \\ 
\rowcolor{gray!10} \cite{liang1986} & GEE & Analyzing repeated measurements of respiratory outcomes, establishing a foundation for their use in public health research & Respiratory Outcomes \\ 
\cite{zeger1988} & GEE & Binary data analysis, focusing on treatment adherence in clinical trials & Clinical Trials \\ 
\rowcolor{gray!10}\cite{neuhaus1991} & GEE & Modeling correlated binary outcomes in healthcare infection rate studies & Infection Control \\ 
\cite{verbeke2000} & LMM & Monitoring HbA1c levels in diabetes trials & Diabetes \\
\rowcolor{gray!10}\cite{esteva2000phase} & LMM & Phase II trial and pharmacokinetic evaluation of cytosine arabinoside for leptomeningeal metastases from breast cancer & Oncology \\ 
\cite{vangeneugden2004applying} & LMM & Estimate the reliability of repeated measurements in clinical trial data for schizophrenia treatment & Psychiatry \\ 
\rowcolor{gray!10}\cite{gueorguieva2004} & LMM & Drug trials evaluating the efficacy of antidepressants across multiple time points & Psychiatry \\ 
\cite{fitzmaurice2008} & LMM & Modeling tumor size reduction over time, handling irregularly spaced follow-up data & Oncology \\ 
\rowcolor{gray!10} \cite{fitzmaurice2008} & GEE & Evaluate weight loss interventions by modeling repeated weight measurements over time & Weight Management \\
\cite{schuler2009temporal} & GEE & Investigated concordance between urine drug screens and self-reported cocaine use over time and across genders & Addiction Medicine \\ 
\rowcolor{gray!10}\cite{chen2013linear} & LMM & Analyzing fMRI group data and study brain activation patterns in response to stimuli & Neuroscience \\ 
\cite{elias2013treatment} & GEE & Analyzed self-efficacy in treatment frameworks among psychology and management scholars & Psychology \\ 
\rowcolor{gray!10} \cite{simoni2013preliminary} & GEE & Conducted a randomized controlled trial of CBT-AD to improve adherence and reduce depression among HIV-positive Latinos & Behavioral Medicine \\ 
\cite{dai2013estimating} & GEE & Estimated the efficacy of preexposure prophylaxis for HIV prevention based on drug concentration thresholds & Epidemiology \\ 
\rowcolor{gray!10}\cite{wiles2014allowing} & LMM & Analysis of non-adherence to treatment in a randomized controlled trial comparing citalopram and reboxetine in treating depression & Psychiatry \\ 
\cite{walker2014models} & LMM & Measuring anthelmintic drug efficacy for parasitologists & Parasitology \\
\rowcolor{gray!10} \cite{cheng2014increases} & GEE & Examined how housing status influenced drug use patterns among street-involved youth in Canada & Public Health \\ 
\cite{song2015linear} & LMM & Modeled QTc prolongation to evaluate the safety profile of olmesartan medoxomil & Cardiovascular Pharmacology \\ 
\rowcolor{gray!10} \cite{halder2017systematic} & GEE & Performed a systematic review of drug efficacy studies for soil-transmitted helminthiases and advocated for individual data-sharing & Parasitology \\ 
\cite{kang2019statistical} & LMM & Statistical analysis of intestinal lesion scores in studies of anti-coccidial drugs in chickens & Veterinary Parasitology \\ 
\rowcolor{gray!10} \cite{guo2019design} & LMM & Design and analysis of mouse clinical trials for oncology drug development & Oncology \\ 
\cite{ketema2021vivo} & LMM & Systematic review and meta-analysis of in vivo efficacy of anti-malarial drugs against clinical Plasmodium vivax malaria in Ethiopia & Infectious Diseases \\ 
\rowcolor{gray!10}\cite{huang2022evaluation} & GEE & Evaluated the effects of skin-to-skin contact on newborn sucking and breastfeeding abilities in a quasi-experimental study & Neonatal Health \\
\cite{mao2025randomized} & LMM & Assessed the efficacy of a digital integrative medicine intervention for cancer patients undergoing treatment & Oncology \\ 
\end{longtable}

\subsubsection{Generalized Estimating Equations}

Generalized Estimating Equations (GEEs), introduced by Liang and Zeger in 1986, are an extension of generalized linear models (GLMs) designed to handle correlated or clustered data, such as repeated measures or longitudinal observations. Unlike LMMs, which explicitly include random effects to account for individual-level variability, GEEs focus on estimating population-averaged effects, making them ideal for studies where the primary interest lies in overall population trends rather than subject-specific inferences. The general form of a GEE is:
\(
g(\mu_{ij}) = X_{ij}\beta,
\)
where \( g(\cdot) \) is the link function (e.g., logit for binary outcomes, log for count data, identity for continuous data), \( \mu_{ij} \) represents the mean response for the \( j \)-th observation of the \( i \)-th cluster, \( X_{ij} \) is the covariate matrix, and \( \beta \) denotes the regression coefficients. GEEs account for within-cluster correlation by specifying a working correlation structure, such as exchangeable, autoregressive, or unstructured. While the correct specification of this structure enhances efficiency, GEEs remain robust even if it is misspecified, providing consistent estimates of regression coefficients.


GEEs are highly flexible, accommodating various outcome types—binary, count, and continuous—via appropriate link functions. They are computationally efficient as they avoid explicit random-effects modeling and are robust to correlation structure misspecification, ensuring reliable population-level inferences. However, they assume data are Missing Completely at Random (MCAR), a stricter condition than the MAR assumption used in LMMs. GEEs focus on population-averaged effects, limiting their ability to provide individual-level inferences or subject-specific predictions. Additionally, their efficiency relies on correctly specifying the working correlation matrix, and they cannot naturally handle complex random-effect structures, restricting their use in hierarchical settings. Despite these limitations, GEEs are widely applied in clinical and epidemiological studies for analyzing correlated data and modeling average treatment effects.

\subsubsection{Modern Extensions of LMMs and GEEs}

Linear Mixed-Effects Models (LMMs) and Generalized Estimating Equations (GEEs) are widely used for analyzing longitudinal and hierarchical data. However, complex relationships often require extensions such as Nonlinear Mixed-Effects Models (NLMMs) and Generalized Additive Mixed Models (GAMMs). NLMMs extend LMMs by incorporating nonlinear functions to model complex biological processes like drug absorption and elimination in pharmacokinetics. These models take the form \(
Y_{ij} = f(X_{ij}, \beta, b_i) + \epsilon_{ij},
\) where \( f(\cdot) \) is a nonlinear function, \( \beta \) represents fixed effects, \( b_i \) accounts for random effects, and \( \epsilon_{ij} \) captures residual errors. Tools such as NONMEM, Monolix, and the R package \texttt{nlme} facilitate the application of NLMMs, although their computational demands are significant \citep{pinheiro2000, davidian2017nonlinear}. GAMMs combine smooth, nonparametric functions with mixed effects to model unknown or varying relationships. The general form of a GAMM is:  
\(
Y_{ij} = X_{ij}\beta + Z_{ij}b_i + \sum_k f_k(X_{ijk}) + \epsilon_{ij},
\)  
where \( \sum_k f_k(X_{ijk}) \) represents smooth functions of predictors, and \( Z_{ij}b_i \) captures the random effects. GAMMs are commonly used in environmental and disease progression studies but may face overfitting and computational challenges in large datasets \citep{wood2017generalized}.

NLMMs and GAMMs expand the flexibility of LMMs and GEEs by accommodating nonlinear and nonparametric relationships, making them invaluable for modeling complex processes in clinical and epidemiological research. However, careful consideration is needed to address computational demands and ensure model interpretability.

\subsubsection{Modeling with Survival Outcomes}
Accelerated Failure Time (AFT) models are commonly used to analyze time-to-event data by directly modeling the logarithm of survival times as a linear function of covariates. The general form of an AFT model is:
\(
\log(T_i) = X_i\beta + \epsilon_i,
\)
where \(T_i\) is the survival time for the \(i\)-th individual, \(X_i\) is the covariate vector, \(\beta\) is the vector of regression coefficients and \(\epsilon_i\) is the error term. AFT models assume that covariates accelerate or decelerate the time to the event by a constant factor. For example, if \(\beta > 0\), the covariate increases survival time by a factor \(e^\beta\); if \(\beta < 0\), it decreases survival time by the same factor. Common distributions for \(\epsilon\) yield specific AFT models, such as the Weibull AFT or log-logistic AFT.

When longitudinal data is integrated with survival response, Joint Models (JMs) are used to address the dependency between longitudinal trajectories and time-to-event processes, offering a unified framework for analyzing such data. The general joint modeling framework consists of two submodels: the longitudinal submodel, given by $Y_i(t) = X_i^\top(t) \beta + Z_i^\top(t) b_i + \epsilon_i(t),$ where \(Y_i(t)\) is the longitudinal outcome for subject \(i\) at time \(t\), \(X_i(t)\) and \(Z_i(t)\) are design matrices for fixed (\(\beta\)) and random (\(b_i\)) effects, respectively, \(\epsilon_i(t) \sim \mathcal{N}(0, \sigma^2)\) represents residual errors, and the survival submodel $h_i(t | M_i(t), w_i) = h_0(t) \exp(w_i^\top \gamma + \alpha M_i(t)),$ where \(h_i(t)\) is the hazard function for subject \(i\) at time \(t\), \(h_0(t)\) is the baseline hazard, \(w_i\) is a vector of baseline covariates with coefficients \(\gamma\), \(M_i(t)\) is a function of the subject-specific longitudinal trajectory, linking the two submodels via the association parameter \(\alpha\). JMs provide more accurate estimates of survival outcomes and dynamic predictions of survival probabilities by incorporating longitudinal biomarker information, thus reducing bias compared to separate analyses. However, they rely on strong parametric assumptions about the underlying processes, which, if misspecified, can lead to biased results. Fitting JMs is computationally intensive, requiring specialized algorithms like the Expectation-Maximization (EM) algorithm or Bayesian Markov Chain Monte Carlo (MCMC) methods. Recent advancements have addressed these challenges by developing efficient algorithms, such as the R package JSM \citep{xu2020semi}, which provides a semiparametric joint modeling framework and offers a maximum likelihood approach to fit these models. 

\begin{longtable}[l]{p{2.5cm}p{2cm}p{8cm}p{2.5cm}}
\caption{Applications of AFT and Joint Models in Clinical Trials} \\
\hline
\rowcolor{gray!20} \textbf{Study} & \textbf{Method} & \textbf{Application} & \textbf{Therapeutic Area} \\
\endfirsthead

\multicolumn{4}{c}%
{{\bfseries Table \thetable\ Continued from previous page}} \\
\hline
\rowcolor{gray!20} \textbf{Study} & \textbf{Method} & \textbf{Application} & \textbf{Therapeutic Area} \\
\endhead

\hline
\endfoot

\hline
\endlastfoot

\cite{farewell1982} & AFT & Analyze long-term survivors in clinical trials, emphasizing applicability in data with short- and long-term survival times & Clinical Trials \\
\rowcolor{gray!10} \cite{wei1992} & AFT & Use of Weibull AFT models to evaluate treatment effects and compare survival times in cancer patients receiving different chemotherapy regimens & Oncology \\
\cite{bradburn2003} & AFT & Comparison of AFT models with proportional hazards models in censored data survival analysis in cancer trials & Oncology \\
\rowcolor{gray!10} \cite{tsiatis2004} & JM & HIV studies to explore relationships between viral load and time to treatment failure & HIV Studies \\
\cite{proustlima2009} & JM & Alzheimer’s disease trials to model the association between cognitive decline and time to dementia onset & Alzheimer’s Disease \\
\rowcolor{gray!10} \cite{rizopoulos2012} & JM & Predict survival probabilities for cancer patients based on tumor progression biomarkers & Oncology \\
\end{longtable}

\tr{While AFT models and the proportional hazards framework remain the most widely used for survival data, several extensions have been proposed to address more complex event structures. When the proportional hazards assumption is violated, alternatives such as time-varying coefficient models, stratified Cox models, Random Survival Forest and Gradient Boosting type methods provide flexible inference by allowing hazard ratios to change over time \citep{hess1994assessing, therneau2000, ghosh2025ensemble}. In studies with complex censoring patterns, including interval-censored, left-truncated, or competing-risk data, specialized likelihood-based and nonparametric estimators have been developed, such as the Fine–Gray subdistribution model for competing risks \citep{fine1999proportional}. Furthermore, when participants may experience multiple or repeated events, recurrent-event models such as the Andersen–Gill counting-process formulation, Prentice–Williams–Peterson (PWP) total- or gap-time models, and Wei–Lin–Weissfeld (WLW) marginal models offer distinct strategies for capturing dependence among event recurrences \citep{oyamada2022comparison}. Together, these approaches expand survival modeling beyond proportional hazards assumptions and accommodate the complexities frequently encountered in modern clinical trials.}

\tr{\subsection{Handling Missing data} Parametric methods such as ANOVA, ANCOVA, and mixed-effects models rely on explicit distributional assumptions and are particularly sensitive to how missing data arise. In most applications, data are assumed to be missing at random (MAR), that is, the probability of missingness depends only on observed quantities. Under this assumption, likelihood-based estimation or restricted maximum likelihood provides valid inference while using all available data \citep{little2019statistical, molenberghs2007missing}. When data are missing not at random (MNAR), for instance, when dropout depends on unobserved outcomes, bias can occur even with correct model specification. In such settings, models that explicitly link the missingness process to the outcome, such as selection, pattern-mixture, or shared-parameter formulations, are required \citep{daniels2008}. Because MNAR assumptions cannot be empirically verified, sensitivity analyses are essential for assessing robustness. Delta-adjustment, tipping-point, and Bayesian prior–based approaches provide interpretable ways to examine how treatment estimates change under alternative assumptions. Clear documentation of the assumed mechanism and sensitivity framework is therefore critical for credible parametric inference.}

\section{NONPARAMETRIC TESTS}
\label{sec:nonparametric}

Nonparametric tests have become essential tools in clinical trials, particularly when the data deviate from the assumptions required by parametric methods, such as normality or homogeneity of variances. \tr{Nonparametric frameworks are particularly useful when the aim is to compare treatment distributions robustly without relying on parametric assumptions, emphasizing inference on ranks or medians rather than means.} These methods are robust and versatile, making them well-suited for analyzing ordinal, skewed, or small-sample data, as well as datasets with outliers. In clinical trials, nonparametric tests are often used to evaluate treatment efficacy in scenarios where the primary interest lies in comparing distributions or ranks rather than means. For example, in trials with repeated measures or longitudinal outcomes, extensions of classical nonparametric tests, such as the Friedman test \citep{friedman1937}, Brunner-Munzel test \citep{brunner2000}, and nparLD \citep{noguchi2012}, provide powerful alternatives to parametric counterparts. Additionally, in studies where only changes from baseline to the final endpoint are available, rank-based methods such as the Wilcoxon signed-rank test or Mann-Whitney U test are frequently applied. These methods are particularly advantageous in early-phase or exploratory trials with limited sample sizes, offering flexibility in handling non-normal or ordinal outcomes. However, despite their robustness, nonparametric tests may exhibit lower power than parametric methods when parametric assumptions are satisfied, and their interpretation can be less straightforward, particularly in complex longitudinal settings. By addressing these challenges and providing distribution-free inference, nonparametric methods play a vital role in efficacy testing, especially in studies involving diverse patient populations or unconventional outcome measures.

\subsection{Cross-Sectional Nonparametric Tests}

While parametric methods such as t-tests and ANOVA are widely used for efficacy testing, they rely on assumptions of normality and homogeneity of variance that may not hold in real-world clinical data. When these assumptions are violated -- such as in the presence of skewed distributions, ordinal outcomes, or small sample sizes -- nonparametric methods provide a robust alternative. The following section explores cross-sectional nonparametric tests, including the Wilcoxon Signed-Rank Test and the Mann-Whitney U Test, highlighting their advantages and applications in clinical trial settings.

\subsubsection{Tests for comparing two groups}
The Wilcoxon Signed-Rank Test \citep{wilcoxon1945} is a non-parametric alternative to the paired t-test, used to compare paired observations when parametric assumptions are violated. It tests whether the median of the differences between paired values is zero by ranking the absolute values of the differences and summing the signed ranks. The test statistic is calculated as \(
W = \sum_{i=1}^{n} R_i \cdot \text{sgn}(D_i),
\) where \(D_i = X_i - Y_i\) and \(\text{sgn}(D_i)\) is the sign of the difference. For small samples, critical values are from exact tables, while for large samples, the statistic approximates a normal distribution. This test is robust for small sample sizes and ordinal data but assumes symmetry in the differences, which if violated, can lead to biased results. It is less powerful than the paired t-test for normally distributed data.

The Mann-Whitney U Test \citep{mann1947} compares two independent groups, assessing if one group tends to have larger values than the other. It is particularly useful for ordinal data or skewed distributions. The U statistic is calculated based on ranks of the pooled data, with the formula:
\(
U = n_1 n_2 + \frac{n_1(n_1+1)}{2} - \sum_{i=1}^{n_1} R(X_i),
\) where \(R(X_i)\) is the rank of \(X_i\) in the combined sample. The test assumes that the groups are independent and that the distributions have the same shape and scale. For small sample sizes, critical values are obtained from exact tables, while larger samples use a normal approximation. It is less efficient than the t-test for normally distributed data, trading power for robustness. Furthermore, it assumes that the groups are independent and the distributions of the two groups have the same shape and scale; otherwise, the results may reflect differences in spread rather than central tendency.

Nonparametric tests for efficacy in a two-group setup has been widely studied in the literature, including \cite{altman1991}, who evaluated blood pressure changes in antihypertensive drug trials, while \cite{gibbons2010} compared pain scores and skin improvement in crossover and dermatology trials, respectively. Other studies assessed the accuracy of diagnostic tools \cite{lehmann2006}, cholesterol level changes in nutritional research \cite{hollander2013}, genetic and biomarker disclosure process \cite{mayan2025genetic} and progression-free survival in oncology \cite{gibbons2010}.

\subsubsection{Tests for comparing more than two groups}
The Kruskal-Wallis (KW) test, developed by Kruskal and Wallis in 1952, is a non-parametric alternative to one-way Analysis of Variance (ANOVA). It is used to compare medians across two or more independent groups, particularly when the assumption of normality or homogeneity of variances is violated. Unlike ANOVA, the Kruskal-Wallis test ranks the data instead of analyzing raw values, making it robust against outliers and non-normal distributions. Consider \(k\) independent groups with sample sizes \(n_1, n_2, \dots, n_k\) and combined total observations \(N = \sum_{i=1}^{k} n_i\). Let \(R_{ij}\) denote the rank of the \(j\)-th observation in group \(i\), where all observations are ranked jointly across groups. The test statistic \(H\) is calculated as \(
H = {12}(N(N+1))^{-1} \sum_{i=1}^{k} ({T_i^2}/{n_i}) - 3(N+1),
\)
where \(T_i = \sum_{j=1}^{n_i} R_{ij}\) is the sum of ranks for the \(i\)-th group. Under the null hypothesis (\(H_0\)), which assumes that all groups are sampled from the same distribution, \(H\) approximately follows a chi-squared distribution with \(k-1\) degrees of freedom for large \(N\). When \(H_0\) is rejected, it indicates significant differences among the group medians. While the test detects differences among groups, it does not identify which specific groups differ. Post-hoc pairwise comparisons must be conducted to pinpoint group differences, often using adjusted rank-based tests or Bonferroni corrections.

\subsubsection{Tests for association between categorical variables}
The Mantel-Haenszel (MH) test, introduced by Mantel and Haenszel in 1959, is a non-parametric statistical method used to evaluate associations between two categorical variables while controlling for a confounding variable. It is widely used in clinical trials and epidemiology to analyze stratified data, particularly in cases where the data are organized into contingency tables across strata. Given \(K\) strata, each represented by a \(2 \times 2\) table, the MH test combines information across all strata to compute a pooled odds ratio and test for homogeneity or association. For each stratum \(k\), the contingency table is represented as:

\[
\begin{array}{|c|c|c|}
\hline
 & \text{Exposed (E)} & \text{Unexposed (U)} \\
\hline
\text{Case (C)} & a_k & b_k \\
\text{Control (Co)} & c_k & d_k \\
\hline
\end{array}
\]

The MH estimate of the common odds ratio (\(\hat{\theta}_{MH}\)), with \(n_k = a_k + b_k + c_k + d_k\) as the total sample size for the \(k\)-th stratum, is calculated as \(
\hat{\theta}_{MH} = {\sum_{k=1}^{K} ({a_k d_k}/{n_k})}/{\sum_{k=1}^{K} ({b_k c_k}/{n_k})}.
\) The Mantel-Haenszel test statistic is computed as:
\(
\chi^2_{MH} = {\big( \sum_{k=1}^{K} (a_k - E_k) \big)^2}/{\sum_{k=1}^{K} V_k},
\)
where \(E_k\) is the expected value of \(a_k\) under the null hypothesis of no association and \(V_k\) is its variance. The test statistic \(\chi^2_{MH}\) follows a chi-squared distribution with 1 degree of freedom under the null hypothesis. A significant result suggests an association between the exposure and the outcome after controlling for the stratifying variable. However, the test assumes that the odds ratio is consistent across strata, which may not always hold in practice. If the homogeneity assumption is violated, the pooled estimate may be misleading. Furthermore, the test is limited to \(2 \times 2\) contingency tables and cannot accommodate complex multilevel or continuous data without extensions. It also requires sufficient sample sizes within each stratum for reliable results.

\begin{longtable}[l]{p{2.5cm}p{2cm}p{7cm}p{4.5cm}}
\caption{Applications of the Kruskal-Wallis (KW) and Mantel-Haenszel (MH) Tests in Clinical Trials} \\
\hline
\rowcolor{gray!20} \textbf{Study} & \textbf{Method} & \textbf{Application} & \textbf{Therapeutic Area} \\
\endfirsthead

\multicolumn{4}{c}%
{{\bfseries Table \thetable\ Continued from previous page}} \\
\hline
\rowcolor{gray!20} \textbf{Study} & \textbf{Method} & \textbf{Application} & \textbf{Therapeutic Area} \\
\endhead

\hline
\endfoot

\hline
\endlastfoot

\cite{furno2000ceftriaxone} & MH & Meta-analysis comparing ceftriaxone with $\beta$-lactams in febrile neutropenia using the Peto-modified MH method to assess relative efficacy & Antibiotic Therapy \\
\rowcolor{gray!10} \cite{luo2012randomized} & KW & Compared the efficacy of various preventive drugs during the course of preventive migraine treatment. & Neurology \\
\cite{tian2013efficacy} & MH & Meta-analysis to assess the efficacy and safety of combining Vandetanib with chemotherapy in advanced non-small cell lung cancer patients & Oncology \\
\rowcolor{gray!10} \cite{de2014efficacy} & MH & Compare clinical success rates of moxifloxacin against pooled active comparator treatments in secondary peritonitis & Antibiotic Therapy \\
\cite{reitsma2014accounting} & MH & Multicenter trials with binary outcomes to summarize data obtained from Early External Cephalic Version (EECV) trials published in 2003 across different strata & Multicenter Trials \\
\rowcolor{gray!10} \cite{takada2015practical} & MH & Infertility treatments clinical trials, emphasizing its simplicity and suitability for crossover designs & Infertility Treatments \\
\cite{chazard2017compare} & KW & Compare the inpatient length of stay (LOS) between different patient samples, exploring the relationship between LOS, treatment benefit, and adverse events. & Healthcare, Inpatient Care \\
\rowcolor{gray!10} \cite{shapiro2019lasmiditan} & MH & Post-hoc analysis evaluating the efficacy of Lasmiditan for treating migraines in patients with cardiovascular risk factors & Neurology \\ 
\cite{President2019} & KW & Explored factors affecting patient participation in clinical research to assess differences in willingness across groups. & Clinical Research \\
\rowcolor{gray!10} \cite{boulware2020randomized} & KW & Evaluated the efficacy of hydroxychloroquine in preventing illness compatible with Covid-19 or confirmed infection when used as postexposure prophylaxis. & Infectious Diseases \\
\cite{hoo2023impact} & MH & Systematic review and meta-analysis on integrated care's impact on outcomes after acute coronary syndrome & Cardiovascular Research \\
\end{longtable}

\subsubsection{Tests with survival data}
The Log-Rank test, first introduced by Mantel in 1966, is a nonparametric statistical method used to compare survival distributions between two or more groups. The test is based on the null hypothesis (\(H_0\)) that there is no difference in the survival distributions between groups. Consider \(k\) groups, and let \(n_i(t)\) and \(d_i(t)\) denote the number of subjects at risk and the number of events observed at time \(t\) in group \(i\), respectively. The observed (\(O_i\)) and expected (\(E_i\)) event counts for each group are calculated as follows:
\(
E_i(t) = ({n_i(t)}/{n(t)}) \cdot d(t),
\)
where \(n(t) = \sum_{i=1}^{k} n_i(t)\) is the total number of subjects at risk at time \(t\), and \(d(t) = \sum_{i=1}^{k} d_i(t)\) is the total number of events at time \(t\). The observed and expected values are then summed across all time points. The test statistic is given by
\(
\chi^2 = {\big( \sum_{i=1}^{k} (O_i - E_i) \big)^2}/{\sum_{i=1}^{k} V_i},
\)
where \(V_i\) is the variance of \(O_i\) under the null hypothesis, calculated as
\(
V_i = (\sum_{t} {n_i(t) \cdot (n(t) - n_i(t)) \cdot d(t) \cdot (n(t) - d(t))})/({n(t)^2 \cdot (n(t) - 1)}).
\) Under \(H_0\), the test statistic \(\chi^2\) follows a chi-squared distribution with \(k-1\) degrees of freedom. For \(k = 2\), the test reduces to a single degree of freedom test comparing two groups. The Log-Rank is a robust, distribution-free method for comparing survival curves and it retains good power when the hazard ratios between groups are proportional. However, violations of the proportional hazards assumption can lead to biased results. The test is also sensitive to censoring patterns; unbalanced censoring across groups can distort the results. Moreover, it does not adjust for covariates, requiring the use of stratified methods for more complex analyses.

Gray’s test, introduced in \cite{gray1988}, is a non-parametric method for comparing cumulative incidence functions in the presence of competing risks. In clinical trials and epidemiological studies, competing risks occur when an individual is at risk of experiencing more than one mutually exclusive event, such as death from different causes. Unlike the Log-Rank test, which assumes all events are of the same type, Gray’s test accounts for the sub-distribution of competing risks. The cumulative incidence function (CIF), denoted as \(F_j(t)\), represents the probability of experiencing event \(j\) by time \(t\), considering the presence of other competing events. Gray’s test assesses whether the CIFs differ significantly between groups for a specific event type \(j\). The test statistic for Gray’s test is derived from weighted differences in observed and expected event counts for each group. For \(k\) groups, let \(d_{ij}(t)\) denote the number of events of type \(j\) at time \(t\) in group \(i\), and \(n_i(t)\) denote the number of individuals at risk in group \(i\) at time \(t\). The weighted observed-minus-expected difference is computed as \(
S(t) = \sum_{i=1}^{k} w(t) \cdot \left[d_{ij}(t) - E_{ij}(t)\right],
\)
where \(E_{ij}(t)\) is the expected number of events in group \(i\) under the null hypothesis, and \(w(t)\) is a weight function. The variance of \(S(t)\) is estimated to compute the test statistic \(
\chi^2 = {S(t)^2}/{\text{Var}[S(t)]}.
\) Under the null hypothesis, \(\chi^2\) follows a chi-squared distribution with \(k-1\) degrees of freedom. This test is found to be particularly valuable in clinical trials where competing risks are prominent, such as cancer studies with multiple causes of mortality. However, Gray’s test is sensitive to the choice of weights, and assumes independence between competing events and censoring, which may not always hold in practice. Additionally, the test does not adjust for covariates, requiring extensions such as Fine and Gray’s regression model for more complex analyses.

\subsection{Longitudinal Nonparametric Tests}
While cross-sectional nonparametric tests provide robust alternatives to parametric methods for single-time-point comparisons, many clinical trials involve repeated measurements over time. In such cases, traditional rank-based methods may fail to account for within-subject correlations, requiring specialized nonparametric techniques for longitudinal data. These methods, such as the Friedman Test, Brunner-Munzel Test, the nparLD framework and the Longitudinal Rank Sum Test, extend nonparametric inference to repeated measures settings, enabling more flexible and assumption-free analysis of treatment effects over time. The following section explores these approaches, their applications, and their advantages in handling non-normal, ordinal, and irregularly spaced longitudinal data.

\subsubsection{Friedman Test}
The \textit{Friedman Test}, introduced by \cite{friedman1937}, is a nonparametric method designed for analyzing repeated measures or matched group data across multiple treatments. The test ranks observations within each subject and evaluates differences in ranks across treatments, making it robust to non-normal distributions. For \( n \) subjects and \( k \) treatments, let \( R_{ij} \) denote the rank of the \( j \)-th treatment for the \( i \)-th subject. The test statistic is computed as:
\(
Q = \frac{12n}{k(k+1)} \sum_{j=1}^k \left( \bar{R}_j - \frac{k+1}{2} \right)^2,
\)
where \( \bar{R}_j = \frac{1}{n} \sum_{i=1}^n R_{ij} \) is the average rank of treatment \( j \). Under the null hypothesis that all treatments have identical distributions, \( Q \) approximately follows a chi-squared distribution with \( k-1 \) degrees of freedom when \( n \) is large. A significant result suggests differences in treatment effects. The Friedman Test is particularly suitable for ordinal data or non-normally distributed outcomes, where traditional parametric methods like repeated measures ANOVA are inappropriate.

The Friedman Test offers simplicity and robustness as its primary advantages. It does not assume normality and is straightforward to implement, making it accessible for small datasets and early-stage studies. However, it comes with notable limitations. The test assumes balanced data, which means every subject must have observations across all time points or treatments---a condition rarely met in real-world longitudinal studies with missing data. Additionally, it treats measurements as independent ranks within subjects, ignoring temporal trends or within-subject variability, which are critical in longitudinal studies. 
Despite these limitations, the Friedman Test remains a foundational method for small-scale repeated measures experiments, especially in fields like nutrition and behavioral studies.

\subsubsection{Brunner-Munzel Test}

The \textit{Brunner-Munzel Test}, introduced by \cite{brunner2000}, is a rank-based nonparametric method designed to compare two groups while allowing for unequal variances and non-normal distributions. Unlike the Wilcoxon Rank-Sum Test, the Brunner-Munzel Test does not assume homogeneity of variances, making it robust in situations where group variances differ. For independent samples \( X_1, \dots, X_n \) from group 1 and \( Y_1, \dots, Y_m \) from group 2, the test statistic evaluates the probability \( P(X < Y) + 0.5 \, P(X = Y) \), interpreted as the stochastic dominance of one group over the other.
The test statistic \( T \) is calculated as:
\(
T = \frac{\hat{\Delta}}{\sqrt{\hat{\sigma}^2}},
\)
where \( \hat{\Delta} \) is the difference in rank-based means between the two groups, and \( \hat{\sigma}^2 \) estimates the variance of the rank means. Under the null hypothesis of no difference between the groups, \( T \) follows a standard normal distribution asymptotically. The Brunner-Munzel Test is particularly useful in comparing treatment effects where variability differs substantially between groups, such as in clinical trials with heterogeneous populations.

The Brunner-Munzel Test offers significant advantages over traditional rank-based methods. Its ability to handle unequal variances makes it particularly robust in practical scenarios, such as comparing treatment efficacy in diverse patient populations or analyzing data with extreme outliers. Additionally, it retains the benefits of nonparametric methods, including robustness to non-normality and suitability for ordinal outcomes. However, the test is limited to two-group comparisons, making it less versatile for studies involving more than two treatments or time points. Moreover, like many rank-based tests, it does not explicitly incorporate temporal dependencies or repeated measures, which restricts its applicability in longitudinal settings. Despite these limitations, the Brunner-Munzel Test remains a valuable tool in scenarios requiring robust, nonparametric two-sample comparisons.


\subsubsection{nparLD Framework}

The \textit{nparLD framework}, introduced by \cite{noguchi2012} is a nonparametric rank-based approach specifically designed for the analysis of longitudinal and repeated measures data. It extends traditional rank-based methods by using pseudo-rank transformations, which preserve the ordinal structure of the data while accounting for repeated measures. For a dataset with \( n \) subjects, \( t \) time points, and \( k \) treatment groups, let \( Y_{ij} \) represent the observation for subject \( i \) at time \( j \). The pseudo-rank \( R_{ij}^* \) of \( Y_{ij} \) is calculated based on its rank relative to all observations across time points and groups. Using these pseudo-ranks, the framework evaluates treatment, time, and interaction effects through F-type test statistics.
The F-type statistic for a given effect is expressed as:
\(
F = {\text{Between-group pseudo-rank variability}}/{\text{Residual pseudo-rank variability}},
\)
where the numerator captures the variability explained by the effect (e.g., group, time, or interaction), and the denominator accounts for residual variability. For example, the test for a group effect compares the average pseudo-ranks across treatment groups, while controlling for time and subject variability. The null hypothesis of no effect is tested using permutation or large-sample asymptotic methods, with the F-statistic approximated by an F-distribution under the null.

The nparLD framework offers several advantages. Its ability to handle unbalanced data and missing observations makes it particularly suitable for real-world longitudinal studies, where dropout and irregular follow-up times are common. Additionally, it accommodates both ordinal and continuous outcomes, providing robust results even in the presence of non-normal data or outliers. The inclusion of interaction effects, such as group-by-time interactions, further enhances its utility for complex study designs. However, there are notable limitations. The need for multiplicity adjustments when testing multiple hypotheses can reduce statistical power, and the computational demands of pseudo-rank transformations and F-statistic calculations increase with larger datasets. While nparLD is robust for moderate-to-large sample sizes, its power may be limited in small-sample settings. Despite these challenges, nparLD has been widely adopted in clinical trials, with applications ranging from the evaluation of repeated measures of biomarkers to studies of vaccine efficacy and weight-loss interventions.

\subsubsection{Longitudinal Rank-Sum Test (LRST)}
The \textit{Longitudinal Rank-Sum Test (LRST)}, introduced by \cite{xu2025novel}, is a nonparametric method developed to evaluate treatment effects across multiple longitudinal endpoints in clinical trials. Unlike traditional rank-based methods, the LRST accounts for the complexity of repeated measures and multi-endpoint designs while maintaining robustness against non-normal distributions and outliers. Let \( x_{itk} \) and \( y_{jtk} \) represent the changes from baseline for subjects in the control and treatment groups, respectively, where \( i \) and \( j \) index subjects, \( t \) denotes time, and \( k \) represents outcomes. Observations are ranked across all subjects, time points, and outcomes, with ranks \( R_{itk} \) and \( R_{jtk} \). The test statistic is calculated as
\(
T_{\text{LRST}} = ({\bar{R}_{y\cdot\cdot\cdot} - \bar{R}_{x\cdot\cdot\cdot}})/{\sqrt{\widehat{\text{Var}}(\bar{R}_{y\cdot\cdot\cdot} - \bar{R}_{x\cdot\cdot\cdot})}},
\)
where \( \bar{R}_{y\cdot\cdot\cdot} \) and \( \bar{R}_{x\cdot\cdot\cdot} \) are the mean ranks for the treatment and control groups, respectively, aggregated across all time points and outcomes. Under the null hypothesis of no treatment effect, \( T_{\text{LRST}} \) asymptotically follows a standard normal distribution. LRST has also been developed for multi-arm clinical trials by \citep{ghosh2025non, ghosh2025power}.

The LRST offers several advantages, making it a valuable tool in modern clinical trials. Its rank-based approach is robust to outliers and non-normal data distributions, which are common in real-world datasets. By simultaneously evaluating multiple longitudinal endpoints, the LRST reduces the need for multiplicity adjustments, enhancing statistical power while maintaining control of Type I error rates. Additionally, it handles missing data and irregular follow-ups efficiently, ensuring flexibility in complex trial designs. However, the test relies on large-sample approximations for its validity, and its performance in small-sample studies may require further evaluation. Despite these limitations, the LRST has shown strong applicability in neurodegenerative disease and oncology trials, where multiple outcomes such as motor function and cognitive performance are monitored over time, providing a comprehensive evaluation of treatment efficacy.

\tr{\subsection{Handling Missing Data:} Nonparametric and rank-based methods are sensitive to missing data because their validity depends on the complete ordering of observations. When data are missing completely at random (MCAR), complete-case analysis remains unbiased but less efficient. Under missing at random (MAR) assumptions, validity can be maintained through several approaches. Inverse Probability Weighting (IPW) reweights observed cases by the inverse of their estimated response probabilities, while Multiple Imputation (MI) replaces missing values with plausible draws from predictive models and combines results across imputations to incorporate uncertainty \citep{schafer1997analysis}. Extensions such as augmented IPW and weighted-rank procedures improve small-sample efficiency and robustness \citep{akritas1997nonparametric, ibrahim2009missing}. 
When data are missing not at random (MNAR), rank-based inference alone cannot guarantee unbiasedness; combining these approaches with targeted sensitivity analyses remains essential for assessing robustness \citep{Li2013SMMR}.}

\begin{longtable}[l]{p{2cm}p{3cm}p{7cm}p{2.5cm}}
\caption{Applications of Nonparametric Methods in Clinical Trials (Sorted by Year)} \\
\hline
\rowcolor{gray!20} \textbf{Study} & \textbf{Method} & \textbf{Application} & \textbf{Therapeutic Area}  \\
\endfirsthead

\multicolumn{4}{c}%
{{\bfseries Table \thetable\ Continued from previous page}} \\
\hline
\rowcolor{gray!20} \textbf{Study} & \textbf{Method} & \textbf{Application} & \textbf{Therapeutic Area} \\
\endhead

\hline
\endfoot

\hline
\endlastfoot

\hline
\rowcolor{gray!10} \cite{bandelow1998use} & Friedman & Panic and agoraphobia scale in a clinical trial & Psychiatry \\
\cite{khan2004systemic} & nparLD & Resistance to Phytophthora crown rot in cucumbers & Agriculture \\
\rowcolor{gray!10} \cite{rorden2007improving} & Brunner-Munzel & Lesion-symptom mapping in cognitive neuroscience & Neuroscience \\
\cite{davis2012effect} & Friedman & Immune cell response to pemetrexed in pancreatic adenocarcinoma & Oncology \\
\rowcolor{gray!10} 
\cite{hofler2016s} & Brunner-Munzel & Ketamine use in refractory status epilepticus & Neurology \\
\cite{tsuboyama2016atg} & Brunner-Munzel & Degradation mechanisms in autophagy systems & Cellular Biology \\
\rowcolor{gray!10} 
\cite{kumar2018stress} & Friedman & Stress-strength estimation in clinical trials & General Medicine \\
\cite{kim2018automatic} & nparLD & Automatic detection of major depressive disorder using electrodermal activity & Mental Health \\
\rowcolor{gray!10} \cite{li2018linkage} & nparLD & Linkage between the I-3 gene for Fusarium wilt resistance and bacterial spot sensitivity in tomato & Agriculture \\
\cite{tuomainen2019human} & Friedman & Compare the efficacy of 19 anticancer compounds on HNSCC cell lines 
& Oncology \\
\rowcolor{gray!10} 
\cite{arenas2019limb} & Friedman & Edetate disodium-based chelation for critical limb ischemia & Diabetes \\
\rowcolor{gray!10} 
\cite{adcock2020effects} & nparLD & Exergame training for older adults & Geriatrics \\
 \cite{teh2020metabolic} & nparLD & Metabolic adaptations to targeted therapies in uveal melanoma & Oncology \\
 \rowcolor{gray!10}
\cite{csenicsik2021effect} & nparLD & Impact of isolation on mental health of athletes during COVID-19 & Mental Health \\
\rowcolor{gray!10} \cite{sirima2022randomized} & Brunner-Munzel & Sporozoite vaccine efficacy for malaria prevention & Infectious Diseases \\
\cite{mashaly2023intralesional} & Friedman & Ethanolamine oleate injection for postoperative pain & Oral Surgery \\
\rowcolor{gray!10} \cite{scharf2023depression} & Brunner-Munzel & Depression and anxiety in ischemic stroke patients & Neurology \\
\cite{yakar2023value} & nparLD & Biomarkers in gingival crevicular fluid during menopause & Periodontology \\
\rowcolor{gray!10} 
\cite{algyar2024laparoscopic} & Friedman & Comparison of pain block methods in bariatric surgery & Anesthesiology \\
\rowcolor{gray!10} 
\cite{mohamed2024efficacy} & Friedman & Epidural prolotherapy versus steroids for chronic pain & Pain Management \\ \hline
\end{longtable}


\section{BAYESIAN MODELS}
\label{sec:bayesian}

Bayesian methods have become indispensable tools for analyzing longitudinal data in clinical trials, offering a probabilistic framework that seamlessly integrates prior knowledge with observed data. By explicitly modeling uncertainty, Bayesian approaches provide a comprehensive understanding of treatment effects, making them particularly well-suited for efficacy testing in complex and high-stakes settings. \tr{Bayesian approaches serve both confirmatory and decision-analytic objectives, integrating prior evidence with observed data to produce interpretable posterior estimates for efficacy or safety outcomes.} These methods excel in handling challenges such as small sample sizes, hierarchical data structures, and irregularly sampled longitudinal data, all of which are common in clinical trials. Despite their numerous advantages, Bayesian methods face challenges, particularly in terms of computational demands and the selection of appropriate priors. However, advancements in scalable inference algorithms, such as variational inference and MCMC methods, continue to mitigate these issues, broadening the scope of Bayesian methods in clinical trials.

\subsection{Bayesian Hierarchical Mixed Model}
Bayesian Hierarchical Models (BHMs) are foundational tools in the analysis of longitudinal clinical trial data. These models account for multi-level data structures, such as repeated measurements nested within subjects, by introducing random effects at different levels of the hierarchy. The Bayesian framework incorporates prior distributions over model parameters, enabling the explicit representation of uncertainty. A typical BHM for longitudinal data can be expressed as
\(
y_{ij} = \beta_0 + \beta_1 t_{ij} + b_i + \epsilon_{ij},
\) where \(y_{ij}\) is the observed outcome for subject \(i\) at time \(j\), \(\beta_0\) and \(\beta_1\) are fixed effects representing the population-level intercept and slope, \(b_i \sim \mathcal{N}(0, \sigma_b^2)\) represents the random effect for subject \(i\), \(\epsilon_{ij} \sim \mathcal{N}(0, \sigma^2)\) is the residual error, and \(t_{ij}\) represents time. In the Bayesian framework, prior distributions are specified for all parameters. For example, \(
\beta_0, \beta_1 \sim \mathcal{N}(0, \tau^2), \quad \sigma_b^2, \sigma^2 \sim \text{Inverse-Gamma}(a, b).
\) The inclusion of prior information allows the incorporation of external knowledge or historical data, further enhancing model robustness. Posterior distributions of the parameters are obtained using MCMC methods or variational inference, enabling uncertainty quantification for all components of the model. However, the reliance on MCMC methods often requires significant computational resources. Furthermore, careful selection of prior distributions is essential, as improper priors may introduce bias into the model estimates. Interpretation of results, while probabilistically robust, can also be challenging in highly hierarchical settings.


\subsection{Bayesian Nonparametrics and Dynamic Bayesian Networks}
Bayesian nonparametric models extend Bayesian methods to settings where the complexity of the model is not fixed a priori but can grow with the data. These models are particularly useful in clinical trials for longitudinal data, where the underlying distributions or clusters may not be well-defined. Bayesian nonparametrics rely on stochastic processes, such as the Dirichlet process (DP) and Gaussian processes (GP), to construct flexible models. A foundational Bayesian nonparametric model is the Dirichlet Process Mixture Model (DPMM), which allows clustering without pre-specifying the number of clusters. The DPMM is represented as \(
G \sim \text{DP}(\alpha, G_0),
\) where \(G\) is the random probability measure, \(\alpha\) is the concentration parameter controlling cluster formation, and \(G_0\) is the base distribution. Observations \(x_i\) are modeled as \(
x_i \sim F(\theta_i), \enskip \theta_i \sim G,
\) where \(F\) is the likelihood function and \(\theta_i\) are parameters drawn from the Dirichlet process. For longitudinal data, extensions like the Hierarchical Dirichlet Process (HDP) and Dependent Dirichlet Process (DDP) have been developed to model repeated measures and temporal dependencies. For example, the DDP allows the distribution \(G_t\) at time \(t\) to evolve over time \(
G_t \sim \text{DP}(\alpha, G_{t-1}),
\) capturing temporal relationships in longitudinal data. Bayesian Additive Regression Trees (BART) is another popular nonparametric model that partitions the data space using an ensemble of regression trees. BART offers a robust way to model complex, nonlinear relationships in longitudinal data.
On the downside, these approaches are computationally intensive, requiring advanced sampling techniques. Furthermore, hyperparameters require careful tuning to ensure meaningful results.

Bayesian Gaussian Processes (GPs) infer distributions over functions, allowing them to adaptively capture the underlying structure in longitudinal trajectories. GPs are particularly well-suited for irregularly sampled data, a common feature in clinical trials. A GP models a set of observations \(y = \{y_1, y_2, \dots, y_n\}\) at corresponding times \(t = \{t_1, t_2, \dots, t_n\}\) as a realization from a multivariate normal distribution \(
y \sim \mathcal{N}(m(t), K(t, t')),
\) where \(m(t)\) is the mean function, often set to zero for simplicity, and \(K(t, t')\) is the covariance function (or kernel) that encodes the similarity between observations at times \(t\) and \(t'\). Common choices for \(K(t, t')\) include the squared exponential kernel \(
K(t, t') = \sigma^2 \exp\big(-{(t - t')^2}/{2\ell^2}\big),
\) where \(\sigma^2\) is the variance and \(\ell\) is the length scale, controlling the smoothness of the function. The Bayesian nature of GPs allows for uncertainty quantification in predictions. Given observed data \((t, y)\), the posterior predictive distribution for a new time \(t^*\) is also Gaussian, that is, \(
p(y^* | t^*, t, y) \sim \mathcal{N}(\mu(t^*), \sigma^2(t^*)),
\) where \(
\mu(t^*) = K(t^*, t)K(t, t)^{-1}y,
\) \(
\sigma^2(t^*) = K(t^*, t^*) - K(t^*, t)K(t, t)^{-1}K(t, t^*).
\) GPs handle irregularly sampled data seamlessly and can accommodate heteroscedastic noise by modifying the covariance structure. But, they face scalability challenges as the computational cost grows cubically with the number of data points due to the inversion of the covariance matrix. Sparse approximations, such as inducing points, have been developed to mitigate this issue but may compromise model accuracy. Additionally, hyperparameter tuning and kernel selection require careful consideration, as these can significantly impact model performance.

\begin{longtable}[l]{p{2.5cm}p{2cm}p{7cm}p{4.5cm}}
\caption{Applications of Bayesian Nonparametric and Graphical Models in Clinical Trials} \\
\hline
\rowcolor{gray!20} \textbf{Study} & \textbf{Method} & \textbf{Application} & \textbf{Therapeutic Area} \\
\endfirsthead

\multicolumn{4}{c}%
{{\bfseries Table \thetable\ Continued from previous page}} \\
\hline
\rowcolor{gray!20} \textbf{Study} & \textbf{Method} & \textbf{Application} & \textbf{Therapeutic Area} \\

\endhead

\hline
\endfoot

\hline
\endlastfoot

\cite{barber2001} & GP & Model tumor growth dynamics in oncology trials, showing their adaptability to complex trajectories & Oncology \\
\rowcolor{gray!10} \cite{whitehead2011bayesian} & BHM & Phase I/II trials investigating the safety and efficacy of drug combinations, focusing on dose finding and exploring therapeutic activity & Dose Finding \\
\cite{ho2013nested} & DP & Nested Dirichlet process model to account for physician-patient interactions in cluster randomized trials, improving treatment effect assessment & Healthcare \\
\rowcolor{gray!10} \cite{koopmeiners2014bayesian} & BHM & Bayesian adaptive design for Phase I/II trials with delayed outcomes, jointly modeling efficacy and toxicity for dose escalation in oncology & Oncology \\
\cite{heinonen2021} & GP & Handle multivariate longitudinal data, enabling simultaneous modeling of multiple biomarkers in cardiovascular trials & Cardiovascular Research \\
\rowcolor{gray!10}\cite{takahashi2021bayesian} & BHM & Bayesian optimization for dose-finding to balance efficacy and toxicity in biologic agent trials & Oncology \\
\cite{ricciardi2023dirichlet} & DP & Regression discontinuity designs to assess treatment effects in clinical trials with thresholds & Clinical Trials Design \\
\rowcolor{gray!10} \cite{park2024personalized} & GP & Identify personalized optimal doses in Phase I/II clinical trials by modeling toxicity and efficacy based on patient biomarkers & Dose Finding \\
\cite{dong2024gaussian} & GP & Optimize nanoparticle formulations for drug delivery, improving encapsulation efficiency and therapeutic efficacy & Drug Delivery \\

\hline
\end{longtable}


\tr{\subsection{Handling Missing Data} A major strength of the Bayesian framework is its coherent treatment of missing data through probabilistic modeling. Missing values are treated as additional unknown parameters and are integrated over during posterior sampling, eliminating the need for ad hoc imputation \citep{schafer1997analysis}. When data are missing at random (MAR), data-augmentation or Gibbs-sampling algorithms naturally propagate uncertainty from incomplete observations into posterior estimates. For missing not at random (MNAR) scenarios, the Bayesian approach allows explicit modeling of the missingness mechanism using selection, pattern-mixture, or shared-parameter formulations \citep{ibrahim2001, daniels2008}. Sensitivity analyses can be incorporated directly into the Bayesian hierarchy by specifying alternative priors for missingness parameters, enabling robust inference under unverifiable assumptions. This unified treatment of uncertainty across both observed and unobserved data makes Bayesian methods especially attractive for longitudinal and adaptive clinical trial analyses.}

\section{DEEP LEARNING BASED METHODS}
\label{sec:deepLearn}

Deep learning has revolutionized data analysis across various domains, including clinical trials, by providing powerful tools to model complex and high-dimensional data. Unlike traditional statistical methods, deep learning models do not rely on predefined assumptions about data distribution or structure. Instead, they leverage neural networks to learn intricate patterns and relationships directly from the data, making them particularly suited for efficacy testing in longitudinal clinical trials with complex, unstructured, or high-dimensional datasets. \tr{Machine learning and deep learning approaches primarily target prediction and adaptive decision-making, using flexible, data-driven algorithms that capture nonlinearities and interactions without explicit distributional assumptions.} Despite their flexibility and power, deep learning methods face challenges in clinical trial settings. These include high computational demands, a reliance on large datasets for training, and difficulties in interpreting model outputs. To address these limitations, recent advancements have focused on integrating deep learning with Bayesian frameworks, enabling uncertainty quantification and enhancing interpretability. Techniques like Bayesian Neural Ordinary Differential Equations (Neural ODEs) and Graph Neural Networks (GNNs) have further pushed the boundaries of deep learning applications in clinical research, enabling continuous-time modeling and the analysis of networked clinical data.

\begin{figure}[htbp]
    \centering
    \includegraphics[width=0.9\linewidth]{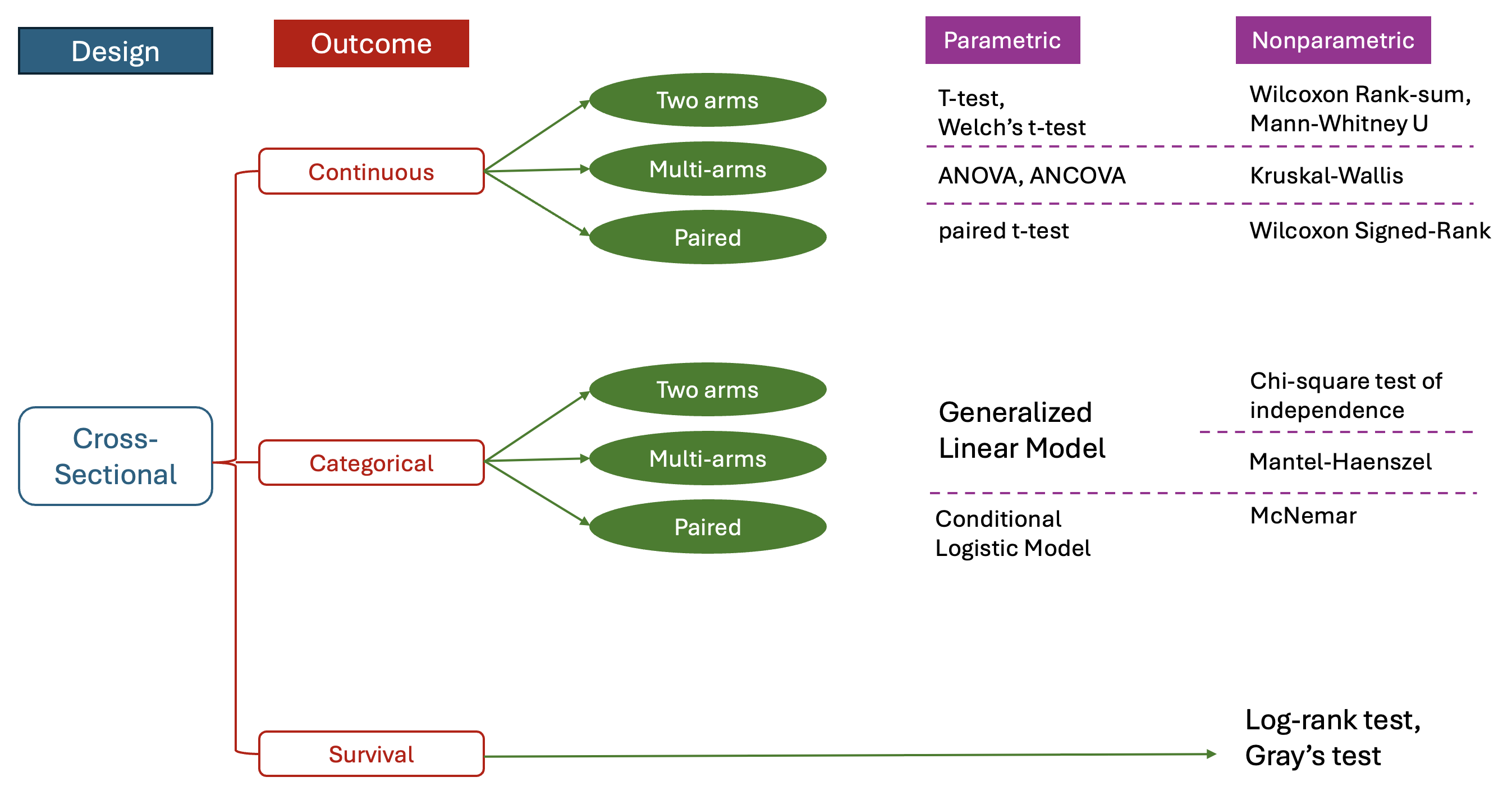}
    \includegraphics[width=0.9\linewidth]{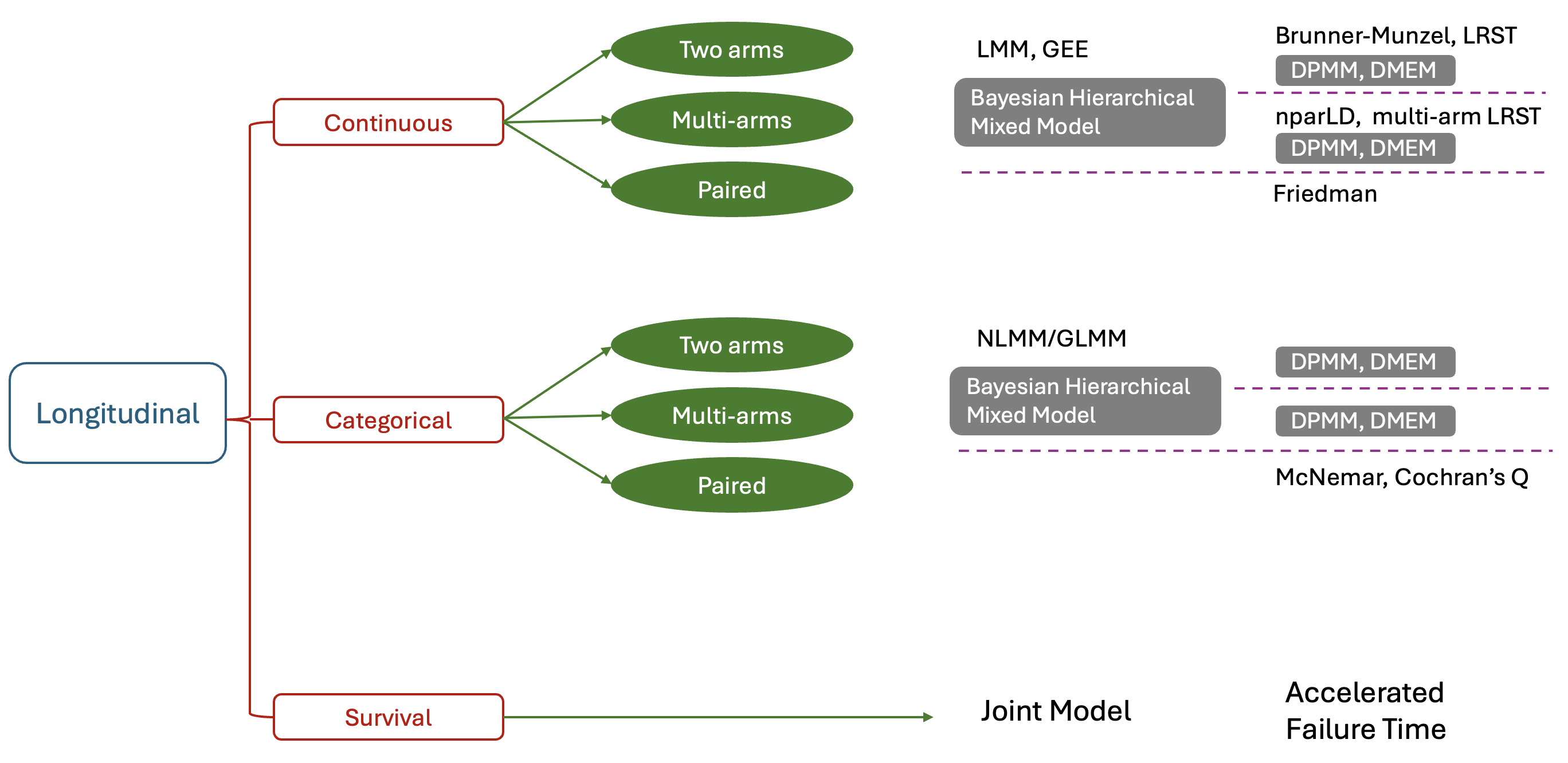}
    \caption{Schematic decision framework linking study design, outcome type, and analytic approach. The diagram guides readers from data structure (cross-sectional or longitudinal) through outcome characteristics (continuous, categorical, survival) to suitable parametric and nonparametric tests for two-arm, multi-arm, or paired comparisons.}
    \label{fig:schematic_diag}
\end{figure}

\subsection{Deep Mixed Effects Models (DMEMs)}
Deep Mixed Effects Models (DMEMs) integrate the hierarchical structure of traditional mixed-effects models with the flexibility of neural networks, enabling the modeling of complex, nonlinear relationships in longitudinal data. Mixed-effects models decompose the observed outcome for subject \(i\) at time \(j\), \(y_{ij}\), into fixed effects (population-level trends), random effects (subject-specific deviations), and residual errors. In DMEMs, this decomposition is enhanced with a neural network \(f(x_{ij}; \theta)\), which captures nonlinear relationships in high-dimensional covariates \(x_{ij}\). The model is expressed as
\(
y_{ij} = f(x_{ij}; \theta) + b_i + \epsilon_{ij},
\)
where \(b_i \sim \mathcal{N}(0, \sigma_b^2)\) represents the random effects specific to subject \(i\), and \(\epsilon_{ij} \sim \mathcal{N}(0, \sigma^2)\) denotes the residual errors. The neural network parameters \(\theta\) are trained alongside random effect variance \(\sigma_b^2\) and residual variance \(\sigma^2\). The neural network \(f(x_{ij}; \theta)\) can take various forms, such as feedforward networks for static data, recurrent neural networks for sequential data, or convolutional networks for spatially structured data. The optimization of DMEMs often involves maximum likelihood estimation (MLE) or Bayesian approaches. DMEMs are particularly powerful for longitudinal data as they allow the estimation of subject-specific trajectories and provide robust predictions even in the presence of missing data or unbalanced study designs. However, DMEMs are computationally demanding, requiring substantial resources to optimize both the neural network and random effects parameters.  The interpretability of
DMEMs can also be challenging due to the black-box nature of neural networks, necessitating the use of feature attribution methods, such as SHAP (SHapley Additive exPlanations)
or LIME (Local Interpretable Model-agnostic Explanations).Moreover, overfitting is a concern, particularly in smaller datasets, and regularization techniques must be carefully implemented.

\subsection{Recurrent Neural Networks and Temporal Convolutional Networks}
These belong to a class of neural networks designed to handle sequential and time-series data. They are particularly well-suited for longitudinal data in clinical trials because they allow for temporal dependencies between observations. Unlike traditional neural networks, recurrent neural networks (RNNs) include recurrent connections that create feedback loops, enabling the network to retain information from previous time steps in a hidden state. This structure makes RNNs powerful tools for modeling dynamic systems where the sequence of observations is crucial. Unlike RNNs, which process data sequentially, temporal convolutional networks (TCNs) leverage convolutional layers to capture temporal dependencies, making them computationally efficient and scalable for long sequences. TCNs are particularly suited for irregular sampling and complex time dependencies often occur.

The basic RNN model for a time-series sequence \(x = \{x_1, x_2, \dots, x_T\}\) computes the hidden state \(h_t\) at time \(t\) as \(
h_t = \sigma(W_h h_{t-1} + W_x x_t + b_h),
\)
where \(W_h\) and \(W_x\) are weight matrices for the hidden state and input, respectively, \(b_h\) is the bias vector, \(\sigma\) is the activation function, typically a hyperbolic tangent (\(\tanh\)) or rectified linear unit (ReLU). The output \(y_t\) at time \(t\) is then computed as
\(
y_t = \phi(W_y h_t + b_y),
\) where \(W_y\) is the output weight matrix, \(b_y\) is the output bias, and \(\phi\) is the output activation function. While RNNs can theoretically model long-term dependencies, in practice, they suffer from the vanishing and exploding gradient problem during training. To address these limitations, two specialized architectures have been developed: Long Short-Term Memory (LSTM) networks and Gated Recurrent Units (GRUs). RNNs, especially LSTMs and GRUs, are powerful for handling sequential data with temporal dependencies. They excel in modeling complex longitudinal relationships and can naturally accommodate varying lengths of time series, making them suitable for clinical trials with irregular follow-ups. However, RNNs are computationally expensive, particularly for long sequences, due to the sequential nature of their computations. Training RNNs is challenging due to issues like vanishing gradients in vanilla RNNs and the need for extensive hyperparameter tuning. LSTMs and GRUs mitigate some of these challenges but at the cost of increased architectural complexity.

The fundamental architecture of TCNs includes the following features, namely, \textbf{Causal Convolutions}, which ensure that predictions at time \(t\) are only influenced by data from \(t\) and earlier, \textbf{Dilated Convolutions}, which introduce gaps between filter applications to expand the receptive field without increasing computational cost, \textbf{Residual Connections} which address vanishing gradient issues and improve gradient flow, and \textbf{Sequence Padding} to maintain the sequence length throughout the network, zero-padding is applied to the input. The output of a TCN is generated by applying successive convolutional layers, culminating in a prediction layer. This architecture makes TCNs highly parallelizable, unlike RNNs, which require sequential processing. Thus, TCNs do not suffer from the vanishing gradient problem inherent in RNNs, ensuring stable training. However, their reliance on fixed kernel sizes and dilation rates may require extensive hyperparameter tuning to achieve optimal performance. Additionally, both these approaches face an interpretability issue due to their black-box nature.

\subsection{Graph Neural Networks (GNNs)}
Graph Neural Networks (GNNs) are deep learning models designed to operate on graph-structured data, where nodes represent entities, and edges encode relationships between them. GNNs are increasingly applied to clinical trials, where data often involves complex relationships between patients, treatments, or time-series features. Unlike traditional deep learning models, which assume data independence, GNNs can learn from interdependent entities, making them well-suited for modeling networks of patients, molecular pathways, or clinical sites. For longitudinal clinical trial data, temporal extensions of GNNs, such as Temporal GNNs or Dynamic GNNs, incorporate time as an additional dimension to model evolving patient relationships or feature dynamics. However, GNNs are computationally expensive, especially for large graphs with dense connections. Training can be challenging due to issues like over-smoothing, where embeddings of all nodes become indistinguishable after several layers. 

\subsection{Federated and Reinforcement Learning with Deep Models}
Federated learning (FL) is a decentralized approach to training deep learning models across multiple data sources without the need to centralize data. This is particularly valuable in clinical trials, where privacy concerns, regulatory constraints, and logistical challenges often prevent data from being pooled across institutions or sites. In federated learning, models are trained locally on each participating site and periodically aggregated to form a global model. Extensions like personalized federated learning allow customization of global models to individual institutions, addressing heterogeneity across sites, without exposing sensitive patient data, ensuring compliance with privacy regulations like HIPAA and GDPR. However, communication overhead between institutions can be substantial, especially with frequent model updates. Data heterogeneity across sites can lead to suboptimal convergence or biased global models. Security concerns, such as model inversion attacks, must also be addressed to ensure participant confidentiality.

Reinforcement Learning (RL) is a framework for sequential decision-making, where an agent learns to maximize cumulative rewards through interactions with an environment. In the context of clinical trials, RL has been employed to optimize treatment strategies by modeling patient responses to interventions as a dynamic process. RL is particularly suited for longitudinal data analysis, where the effects of treatments unfold over time. RL offers the ability to personalize treatment strategies dynamically, adjusting decisions based on patient responses over time. However, RL requires substantial amounts of data to train effectively, which can be a challenge in clinical settings with limited patient samples.

\begin{longtable}[c]{p{2.5cm}p{2cm}p{7cm}p{4.5cm}}
\caption{Applications of Deep Learning Methods in Clinical Trials. RL, Reinforcement Learning; TCN, Temporal Convolution Network; GNN, Graph Neural Network; RNN, Recurrent Neural Network; FL, Federated Learning; DMEM, Deep Mixed Effects Model.} \\
\hline
\rowcolor{gray!20} \textbf{Study} & \textbf{Method} & \textbf{Application} & \textbf{Therapeutic Area} \\
\endfirsthead

\multicolumn{4}{c}%
{{\bfseries Table \thetable\ Continued from previous page}} \\
\hline
\rowcolor{gray!20} \textbf{Study} & \textbf{Method} & \textbf{Application} & \textbf{Therapeutic Area} \\
\endhead
\hline
\endfoot
\hline
\endlastfoot

\cite{zhao2009reinforcement,zhao2011reinforcement} & RL & Cancer clinical trial design & Oncology \\
\rowcolor{gray!10} \cite{komorowski2018} & RL & Sepsis management, recommending dynamic treatment adjustments based on evolving patient conditions & Critical Care \\
\cite{harutyunyan2019} & RNN & Use of multitask learning for mortality and length-of-stay prediction in critical care settings & Critical Care \\
\cite{covert2019temporal} & TCN & Seizure detection using temporal graph CNNs & Neurology \\
\rowcolor{gray!10} \cite{kok2020automated} & TCN & Sepsis prediction using TCNs & Critical Care \\
\cite{sheller2020} & FL & Develop machine learning models for brain tumor segmentation with multi-center imaging data, preserving patient privacy & Oncology \\
 \rowcolor{gray!10}
\cite{dayan2021federated} & FL & Predicting COVID-19 outcomes using federated models & Infectious Diseases \\
\rowcolor{gray!10} \cite{nath2022reinforcement} & RL & Potential applications of RL in ophthalmology & Ophthalmology \\
\cite{casy2022assessing} & RNN & Estimation of Jadad’s score for clinical trial robustness & Clinical Trial Methodology \\
\rowcolor{gray!10} \cite{hong2023prediction} & DMEM & Mixture model for healthcare time-series data with Gaussian processes & Healthcare Analytics \\
\cite{zhu2023prediction} & RNN & Chronic kidney disease progression prediction using EHRs & Nephrology \\
\rowcolor{gray!10} \cite{trella2024deployed} & RL & Online RL in oral health clinical trials & Oral Health \\
\rowcolor{gray!10} \cite{xu2024deep} & DMEM & Real-time monitoring in additive manufacturing with mixed-effects models & Manufacturing Processes \\
 \cite{pan2024personalized} & DMEM & Personalized prediction of Parkinson’s progression using Gaussian processes & Neurology \\
\end{longtable}

\tr{\subsection{Handling Missing Data} Missing data are pervasive in clinical trial settings, particularly in multimodal and longitudinal studies that integrate imaging, omics, and clinical outcomes. Deep learning models, while highly flexible, generally require complete input tensors, making missingness a key challenge for model reliability and inference. Traditional solutions rely on multiple imputation or inverse probability weighting (IPW) before model fitting, but recent approaches embed missingness handling directly within the network architecture. Generative models such as Variational Autoencoders (VAEs) and Generative Adversarial Imputation Networks (GAIN) reconstruct missing values by learning joint latent representations of observed and unobserved features, thus preserving biological structure and temporal coherence \citep{yoon2018gain, camino2019improving}. For time-dependent data, Recurrent Neural Networks and Temporal Convolutional Networks have been extended to incorporate missing-indicator embeddings or time-gap encodings, enabling dynamic imputation during sequence learning \citep{che2018recurrent, cao2018brits}. Despite these advances, most deep models assume data are missing at random (MAR); explicitly modeling missing not at random (MNAR) mechanisms remains an open challenge. Integrating deep generative frameworks with Bayesian or causal formulations provides a promising direction for handling informative dropout and achieving principled uncertainty quantification in high-dimensional clinical trial data.}

\section{Conclusion and Future Directions}


\tr{A conceptual integration framework summarizing the links between study design, endpoint type, and analytic approach is presented in Figure \ref{fig:schematic_diag} to provide readers with a practical overview of the decision pathways discussed throughout this review. This schematic highlights how the choice of statistical test naturally follows from the structure of the data and the nature of the outcome, serving as a bridge between traditional inferential methods and emerging data complexities. Building on this framework, several challenges and opportunities remain that warrant further exploration. The growing complexity of clinical data, driven by the integration of multi-modal biomarkers, electronic health records, and real-world evidence, demands novel statistical and computational approaches that can handle high-dimensional, heterogeneous, and often incomplete datasets.}

One key area of future work is the development of hybrid models that combine the strengths of parametric and nonparametric methods. While parametric methods offer efficiency and interpretability under ideal assumptions, nonparametric approaches provide robustness to deviations from these assumptions. Hybrid frameworks could leverage both perspectives, providing flexible yet interpretable solutions for efficacy testing. For example, the integration of rank-based methods with mixed-effects modeling could offer a promising avenue for analyzing complex longitudinal data.

Another promising direction is the advancement of Bayesian methodologies for efficacy testing. Although Bayesian methods have gained traction for their ability to incorporate prior information and quantify uncertainty, their application to high-dimensional and dynamic datasets is still evolving. The development of computationally efficient Bayesian frameworks, such as scalable MCMC algorithms or variational inference techniques, could enable their broader adoption in large-scale clinical trials.

In the realm of machine learning, deep learning models and reinforcement learning strategies have demonstrated potential in optimizing trial designs, predicting outcomes, and personalizing treatments. However, their utility in efficacy testing remains underexplored. Challenges such as lack of interpretability, risk of overfitting, and the need for large labeled datasets present significant barriers to their widespread application. Future research should focus on developing interpretable and domain-specific machine learning models tailored to clinical trial data, particularly those involving time-dependent outcomes.
Moreover, the field lacks a consensus on the best practices for handling missing data in efficacy testing. While methods such as multiple imputation and mixed-effects modeling are widely used, their assumptions and limitations can vary significantly across trials. Further work is needed to develop robust, assumption-free approaches for handling missingness, particularly in longitudinal and adaptive trial designs.


\tr{Ensuring fairness in clinical efficacy testing has become increasingly critical as modern trials incorporate diverse patient populations and complex data sources. Statistical and algorithmic frameworks must therefore guard against biases that can arise from unequal subgroup representation, differential data quality, or model overfitting to majority cohorts. Recent strategies include stratified design and randomization, covariate adjustment for underrepresented groups, and reweighting or balancing techniques that equalize subgroup influence during estimation and prediction. In data-driven frameworks such as machine learning and deep learning, fairness-aware loss functions, adversarial debiasing, and post-hoc calibration methods have shown promise for improving equitable model performance across demographic groups. Integrating these fairness principles into efficacy analyses enhances the generalizability and ethical robustness of statistical inference, ensuring that emerging clinical evidence benefits all populations equitably.}

\tr{Looking forward, several emerging research priorities are poised to shape the next generation of clinical efficacy methodology. One key direction involves developing robust and computationally efficient models for high-dimensional longitudinal data that integrate multimodal inputs such as imaging, genomics, and digital health records while preserving statistical validity \citep{diggle2002analysis, verbeke2000}. Another important challenge is balancing predictive accuracy with causal interpretability, as machine learning and deep learning frameworks become central to trial analytics \citep{doshi2017towards, blei2017variational}. Integrative approaches that couple traditional inferential rigor with scalable representation learning, such as hybrid Bayesian–machine learning models, offer promising pathways toward interpretable yet flexible inference. The rapid rise of virtual and decentralized clinical trials (VCTs), enabled by telehealth, wearables, and remote monitoring, introduces new opportunities and challenges related to irregular data streams, device-based measurement error, and adherence variability \citep{abadi2020virtual,tushar2025virtual, dahal2025xcat, dorsey2016state, izmailova2021remote}. Finally, ensuring fairness, transparency, and transportability in these data-rich settings is essential for generating equitable and generalizable evidence \citep{subbaswamy2020development, chen2013linear}. Together, these directions underscore a shift toward methodological frameworks that are statistically rigorous, computationally adaptive, and ethically grounded, advancing the science of clinical evaluation in an era of data-driven and personalized medicine.}

In summary, while the field has seen substantial progress, opportunities remain to address gaps in integrating advanced computational techniques, improving scalability and interpretability, and ensuring equitable and inclusive efficacy testing. Bridging these gaps will not only enhance the reliability of clinical trial outcomes but also pave the way for more personalized and effective healthcare solutions. By addressing these challenges, future research can ensure that statistical methodologies continue to evolve in step with the increasing complexity and scope of modern clinical trials.

\bibliographystyle{unsrt}  
\bibliography{references}

\end{document}